\def\year{2021}\relax
\documentclass[letterpaper]{article} 
\usepackage{aaai21}  
\usepackage{times}  
\usepackage{helvet} 
\usepackage{courier}  
\usepackage[hyphens]{url}  
\usepackage{graphicx} 
\usepackage{natbib}  
\usepackage{caption} 
\usepackage{xcolor}
\usepackage{amsmath,amssymb,amsfonts}
\usepackage{subcaption}
\usepackage{color, colortbl}
\definecolor{Gray}{gray}{0.9}   
\graphicspath{{./figs/}}
\usepackage{amsthm}
\theoremstyle{definition}
\newtheorem{crp}{Definition}
\newtheorem{genslot}[crp]{Definition}
\newtheorem{virtual}[crp]{Definition}

\newtheorem{flow_v}[crp]{Definition}
\newtheorem{diff}[crp]{Definition}

\theoremstyle{corollary}
\newtheorem{corr1}{Corollary}

\theoremstyle{lemma}
\newtheorem{lem1}{Lemma}
\newtheorem{lem2}[lem1]{Lemma}

\newtheorem{theo1}{Theorem}

\title{Distributed Fair Scheduling for Information Exchange in Multi-Agent Systems}
\author {
    Majid Raeis,\textsuperscript{\rm 1}
    S. Jamaloddin Golestani \textsuperscript{\rm 2} \\
}
\affiliations {
    \textsuperscript{\rm 1} University of Toronto, Toronto, Canada \\
    \textsuperscript{\rm 2} Sharif University of Technology, Tehran, Iran \\
    m.raeis@mail.utoronto.ca, golestani@ee.sharif.edu,
}
\begin{document}

\maketitle
\begin{abstract}
Information exchange is a crucial component of many real-world multi-agent systems.
However, the communication between the agents involves two major challenges: the limited bandwidth, and the shared communication medium between the agents, which restricts the number of agents that can simultaneously exchange information. While both of these issues need to be addressed in practice, the impact of the latter problem on the performance of the multi-agent systems has often been neglected. This becomes even more important when the agents' information or observations have different importance, in which case the agents require different priorities for accessing the medium and sharing their information. Representing the agents' priorities by fairness weights and normalizing each agent's share by the assigned fairness weight, the goal can be expressed as equalizing the agents' normalized shares of the communication medium. To achieve this goal, we adopt a queueing theoretic approach and propose a distributed fair scheduling algorithm for providing weighted fairness in single-hop networks. Our proposed algorithm guarantees an upper-bound on the normalized share disparity among any pair of agents. This can particularly improve the short-term fairness, which is important in real-time applications. Moreover, our scheduling algorithm adjusts itself dynamically to achieve a high throughput at the same time. The simulation results validate our claims and comparisons with the existing methods show our algorithm's superiority in providing short-term fairness, while achieving a high throughput.
\end{abstract}

\section{Introduction}
Many real-world multi-agent systems rely on the information exchange between the agents. However, the communication of the agents involves major practical limitations such as the shared communication medium and the limited bandwidth.
Whether the information exchange is required for the coordination of agents, or solely the transfer of information from one point to another, the agents cannot transmit their messages simultaneously over the same communication medium. 
This becomes even more challenging when there is no coordinator to arbitrate access to the medium and therefore, a distributed algorithm is required for scheduling the agents' transmissions. Fairness is a key concern in distributed scheduling algorithms, where some agents might hog the communication medium for a long period of time and therefore, lead to starvation of the other agents. Moreover, the agents' messages might have different importance or different latency requirements. Information exchange in multi-agent reinforcement learning (MARL) is one such example, where the importance of each agent’s partially observed information can be different from one another~\cite{kim}. Weighted fair queueing, which has long been used as an effective basis for resource allocation and scheduling, is a natural candidate for dealing with these cases. 
The main goal of the weighted fair queueing algorithms is to provide service in proportion to some specified service shares (also known as fairness weights) to the competing users of a shared resource. The Generalized Processor Sharing (GPS) scheme \cite{GPS} serves as the reference for most of the existing fair queueing algorithms in the literature. The GPS algorithm uses a fluid flow model in which all the users can receive service simultaneously. However, because of the fluid flow assumption, the GPS scheme is not applicable in many real-world applications such as information exchange.
Consequently, alternative fair queueing algorithms such as \cite{SCFQ,SFQ} have been proposed to mimic the behaviour of the GPS algorithm in real-world packet-based applications. However, the majority of these algorithms are centralized in implementation and cannot be used directly to provide fair queueing in a distributed manner. 

In the information exchange context, the communication medium is the shared resource that needs to be scheduled among the agents. 
In wireless networks, the medium access control (MAC) protocol is responsible for scheduling the users. IEEE 802.11 DCF\footnote{We use the terms 802.11 DCF and Wi-Fi interchangeably for referring to the Distributed Coordination Function (DCF) protocol in IEEE 802.11} (Wi-Fi) is the dominant distributed MAC protocol in wireless networks.
Although there has been some effort in designing distributed fair scheduling algorithms based on 802.11 DCF, almost all these proposed methods take heuristic approaches to emulate a particular centralized fair queueing algorithm without providing any guarantees or theoretical analysis. Moreover, they suffer from short-term unfairness, which is particularly important in real-time applications. In contrast, we propose a distributed fair scheduling algorithm, which provides deterministic guarantees on the normalized share disparity among any pair of agents. 
Specifically, we propose a distributed scheduling algorithm for information exchange in single-hop wireless networks, which guarantees a bounded disparity between the normalized services received by any two agents. To the best of our knowledge, this is the first time that a distributed scheduling algorithm is capable of providing a bounded service disparity among the users. This particularly improves the short-term fairness of our proposed algorithm compared to the existing ones. Furthermore, our algorithm dynamically adjusts itself to balance the trade-off between the fairness and the network throughput.
\subsection{A Brief Background on 802.11 Wi-Fi}
Wi-Fi is the dominant protocol for wireless communication scheduling, which is based on CSMA (Carrier-Sense Multiple Access). The main idea of CSMA is that each agent is required to listen to the medium before its transmission. The agent is allowed to transmit its message only if the medium is idle. However, to avoid collisions after busy periods, it uses the Binary Exponential Backoff (BEB) mechanism, which requires the agents to choose random backoffs in the interval $[0, CW]$. Whenever the medium becomes idle, each agent discretizes the time into time slots of predefined length (denoted by $\sigma$) and decrements its backoff counter by one after each idle time slot. The backoff counter will be frozen during the busy periods. When the counter reaches zero, the agent is allowed to attempt a transmission.
The parameter $CW$ is called the contention window, which is doubled by the agent if its transmission is unsuccessful. $CW$ is reset to its default value once the agent transmits its message successfully. Moreover, 802.11 uses interframe spaces (IFS), which are specified waiting periods between transmission of frames, to prioritize different traffic types (such as control and data). 
\subsection{Related Work}\label{lit}

Efficient and fair communication between agents of a system is a challenging task. Particularly, it is a key component in applications that require agents' coordination, such as cooperative multi-agent reinforcement learning. Most of the existing studies on multi-agent communication only focus on the bandwidth limitation of the medium and ignore the shared medium contention between the agents~\cite{goldman,mao,foerster,zhang}. For instance, a message pruning mechanism is proposed by \cite{mao} to improve the bandwidth efficiency of the recent deep RL-based communication methods for multi-agent systems~\cite{foerster, jiang, peng, singh}. In another work, 
\cite{zhang_var} introduce Variance Based Control (VBC) technique for efficient communication in MARL, which limits the variance of the exchanged messages between the agents in the training phase. \cite{goldman} have developed a theoretical model for decentralized control of  a cooperative multi-agent system with communication. This model can be used for studying the trade-off between the value and cost (including bandwidth cost) of the exchanged information, as well as its effect on the joint utility of the agents.   While almost all these works focus on the bandwidth efficiency of the communication and ignore the shared medium contention between the agents, \cite{kim} study the problem of communication scheduling in MARL. The authors propose SchedNet for scheduling the communication of the agents, which uses a weight-based scheduling algorithm to prioritize the agents with more important information.

On the other hand, the scheduling problem has been well studied in the area of wireless communication. Since Wi-Fi is the dominant distributed MAC protocol in wireless networks, most of the existing distributed scheduling algorithms have been proposed based on this protocol. In particular, these algorithms use one of the parameters of the 802.11, such as $CW$size~\cite{PMAC}, IFS~\cite{DDRR,IDFQ} or backoff intervals~\cite{DFS}, to emulate centralized fair queueing service disciplines in a distributed manner.
A common problem with these algorithms is the short-term unfairness, which is mainly caused by the random backoffs that are used in the collision resolution mechanism of 802.11~\cite{choi}. In order to improve the short-term fairness,~\cite{pdfs} propose a protocol that uses pulse transmissions for giving priority to the nodes that have experienced collision in their last attempt.  Although this protocol improves the short-term fairness, the order in which packets are served could  still drastically change from the reference centralized fair queueing algorithm. Moreover, almost all these methods use heuristic methods to mimic the behaviour of the centralized fair queueing algorithms and do not provide any guarantees or theoretical backings on the fairness or the network throughput.  


\section{System Model and Problem Formulation}\label{model}

\begin{figure}[!t] 
\centering
\subcaptionbox{}{\includegraphics[width=0.48\columnwidth]{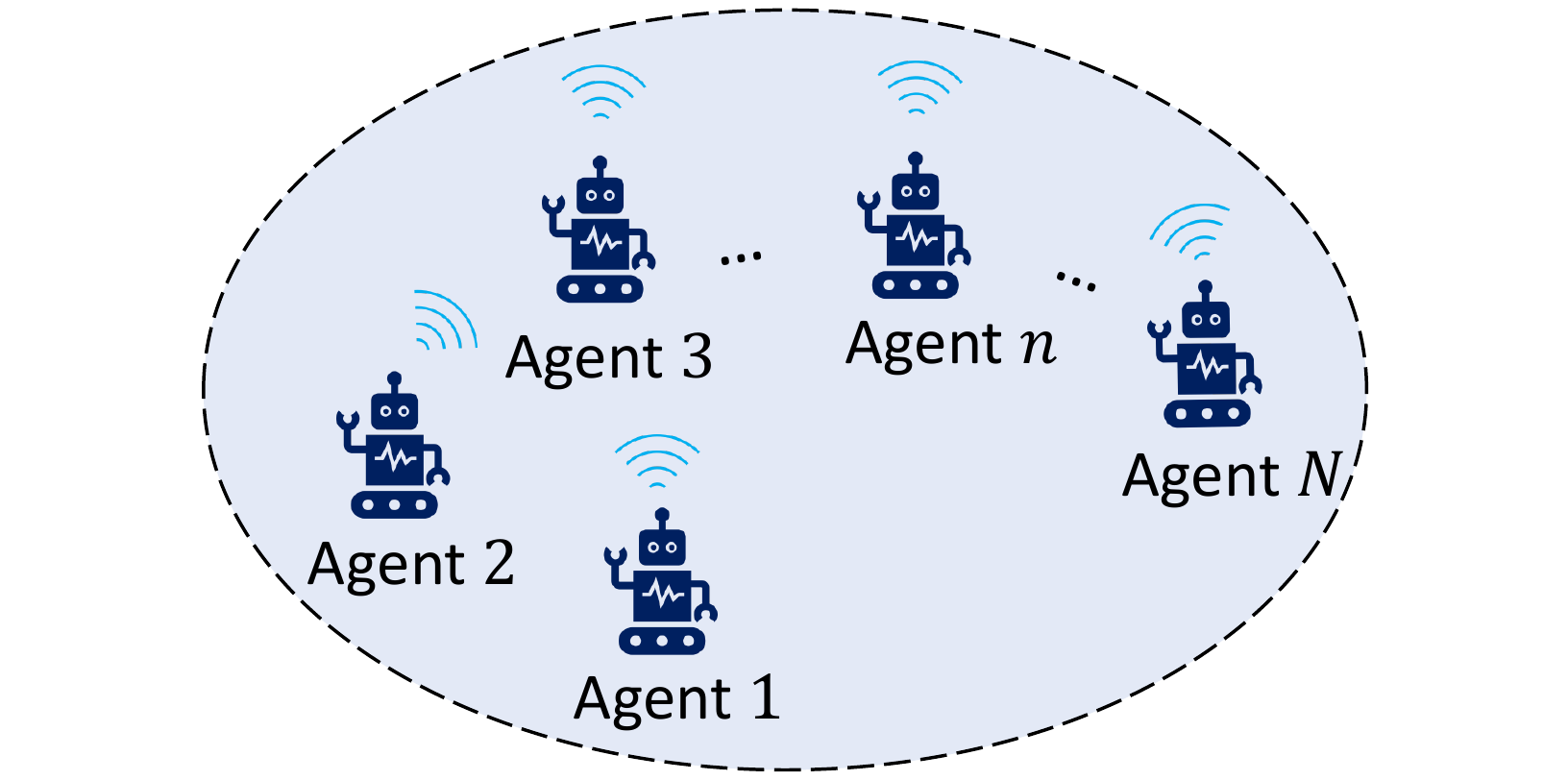}}
\subcaptionbox{}{\includegraphics[scale=0.33]{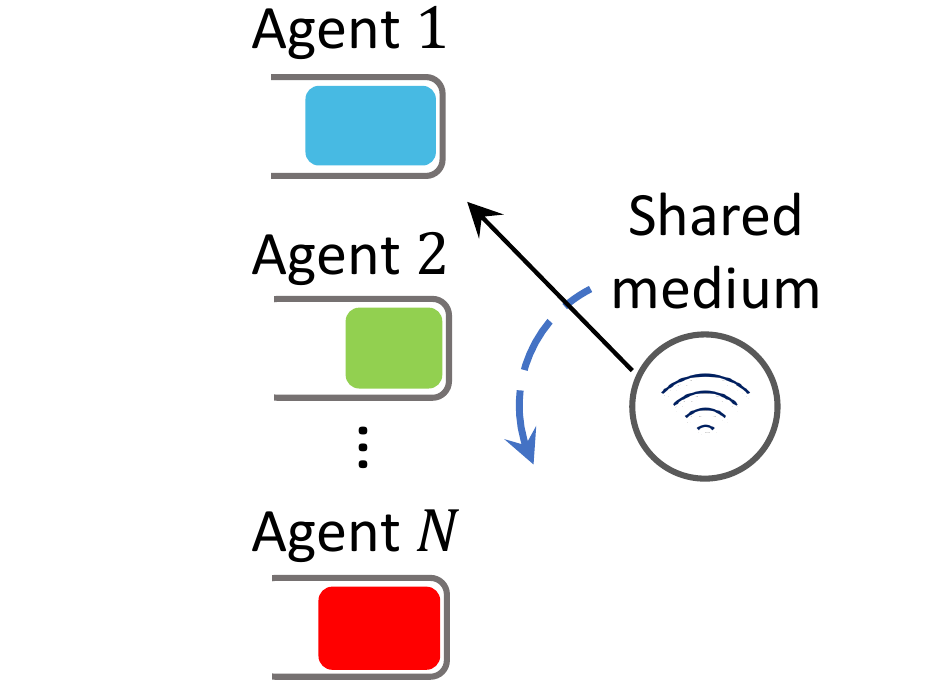}}
\caption{Modeling the communication medium as a shared server in a queueing model that needs to be scheduled between the agents}
\label{fig:model}
\end{figure}

Consider a network consisting of $N$ agents that need to communicate with each other to achieve a common goal. The set of all agents is represented by $\mathcal {N}$ and each agent $n \in  \mathcal {N}$ has a fairness weight of $\phi_{n}$, which corresponds to its share of the communication medium. Fairness weights are chosen independently and are based on the importance of each agent's information. We assume that the agents can sense the medium and therefore, avoid transmitting their messages when the medium is busy. Without loss of generality and solely to simplify the discussion, we assume that all the agents can instantaneously detect if the communication medium becomes busy or idle. Although in practice the agents are only able to detect the medium's status with some delay, the concept of interframe spaces in the 802.11 DCF protocol can be easily merged into our algorithm to address this issue for real-world applications. 


In order to study this problem, we model the communication medium as a shared resource (server) that needs to be scheduled among the agents. Specifically, each agent has a transmission queue for storing the information that needs to be sent (Fig.~\ref{fig:model}). At a given time, the agents with non-empty queues are called \emph{backlogged agents}, while the ones with no information to share are called \emph{absent} agents. 
Let $\mathcal{B}(t)$ denote the set of backlogged agents at time $t$, and $\mathcal{A}(t)$ represent the absent agents. In the same way, $\mathcal{B}(t_{1},t_{2})$ and $\mathcal{A}(t_{1},t_{2})$ respectively represent the set of agents that are backlogged and absent during the whole interval $(t_{1},t_{2})$. Each time that an agent successfully transmits a packet, it receives a service equal to the transmitted packet size. 
We define $W_{k}(t)$ as the total service provided to agent $k$ during $(0,t)$. Based on our definition, if a packet of agent $k$ is under transmission at $t$, $W_{k}(t)$ includes any part of it transmitted by time $t$. The \emph{normalized service} received by agent $k$, $w_{k}(t)$,  is defined as the aggregate service provided to this agent during $(0,t)$ normalized to its fairness weight, i.e., $ w_{k}(t)~=~W_{k}(t)/\phi_{k}, k~\in~\mathcal{N}$. Furthermore, we define bivariate process $w_{k}(t_{1},t_{2})$ as $w_{k}(t_{1},t_{2})~=~w_{k}(t_{2}) -w_{k}(t_{1}), k \in \mathcal{N}$.
 
Now, our goal can be formulated as designing a distributed scheduling algorithm that guarantees a bounded service disparity among any pair of backlogged agents, i.e.,
\begin{equation}
|w_k(t_1,t_2)-w_j(t_1,t_2)|\leq \varepsilon, \qquad k,j  \in ~ \mathcal{B}(t_1,t_2),\label{guarantee}
\end{equation}
while achieving a high throughput at the same time. Since our proposed algorithm implements the centralized \emph{Self-Clocked Fair Queueing} algorithm~\cite{SCFQ} in a distributed manner to provide the above guarantee, we refer to it as Distributed SCFQ (DSCFQ) in the rest of the paper.

\begin{figure}[!t]
\centering
\includegraphics[scale=.35]{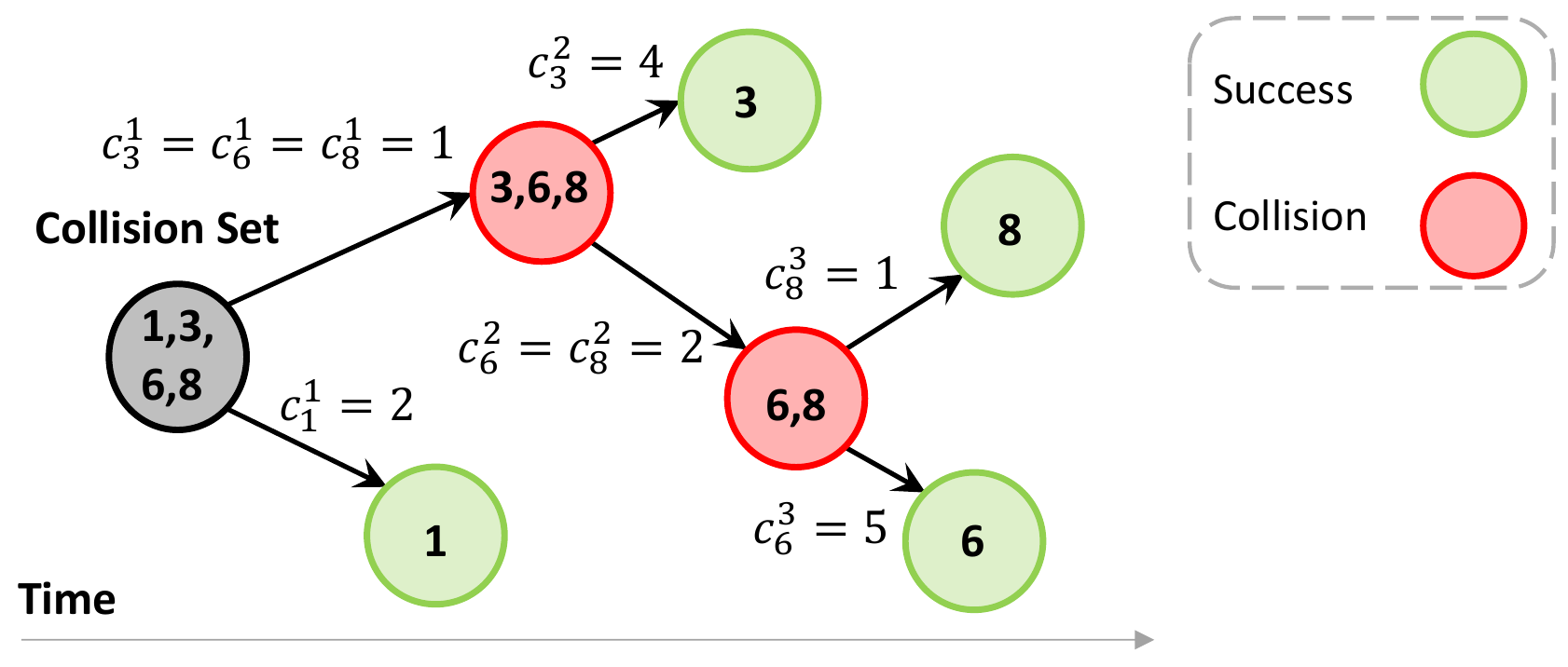}
\caption{ An example of using splitting algorithm (with m = 2) for contention resolution. }
\label{fig:splitting}
\end{figure}

\section{Distributed SCFQ (DSCFQ)}\label{dscfq_algo}

An important source of short-term unfairness in the existing scheduling schemes is that the agents with unsuccessful last attempts have no priority over other agents.
In order to address this issue, we split the competing nodes into two classes: class~I that is defined as the set of agents that have to retransmit their messages because of their unsuccessful last attempts, and class~II nodes, which are the agents with new information to share. 
Moreover, each agent maintains a \emph{collision counter}, \emph{scaling factor} and a \emph{compensation factor}. Collision counter keeps track of the number of collisions that the agent has experienced since its last successful transmission. Scaling factor is a  parameter for suitable scaling of the backoff intervals, in order to reach an acceptable throughput. Finally, compensation factor is a parameter that is required for providing short-term fairness, which compensates the fairness deviations that are caused by the discrete backoff intervals.
 
Now, let us introduce our distributed scheduling algorithm for providing weighted fairness in a multi-agent system.  Let $p_k^i$ denote the $i$th message\footnote{Throughout the paper, we use the terms message and packet interchangeably.} (packet) of agent $k$. Moreover, the arrival time, transmission start time and transmission finish time of packet $p_k^i$ are denoted by $a_k^i$, $s_k^i$ and $d_k^i$, respectively. Now, each agent takes the following actions to transmit the backlogged messages in its queue.

\noindent \textbf{Class~II agents:}
\begin{enumerate}
\item At the time message $p_k^i$ reaches the front of agent $k$'s transmitter queue, it is tagged with backoff tag $B_{k}^{i}$, calculated as follows:
\begin{equation}
B_k^i=\bigg\lfloor \alpha \big(\frac{L_k^i}{\phi_k} - \epsilon_k^i \big) \bigg\rfloor, \label{bkoff}
\end{equation}

where $L_k^i$ and $\alpha$  are the size of packet  $p_k^i$ and scaling factor, respectively, and $\epsilon_k^i $ represents the  compensation factor, which is defined as
\begin{equation}
\epsilon_k^1 = 0, \qquad
\epsilon_k^i  = \epsilon_k^{i-1} + \big(\frac{B_k^{i-1}}{\alpha} - \frac{L_k^{i-1}}{\phi_k}\big). \label{eq:epsilon_def}
\end{equation}
\item To send $p_k^i$, agent $k$ picks a backoff interval equal to  $B_k^i$ slots. Class~II agents can start decrementing their backoff counters only after sensing the medium idle for one time slot. The backoff counter is decremented by one after each idle slot and is frozen during the busy periods.
\item Whenever a node's backoff reaches zero, it starts the transmission process for the intended destination.
\item If a transmitting class~II agent detects a collision, it stops the transmission, increments its collision counter by 1 and starts following the procedure for class~I nodes.
\end{enumerate}
\noindent \textbf{Class~I agents:}
\begin{enumerate}
\item \textbf{Splitting Algorithm for Collision Resolution:} In contrast to class~II agents, each class~I node can access the medium as soon as it becomes idle, without waiting for an extra idle time slot. In this way, the nodes involved in the collision get prior access to the medium. However, to avoid a guaranteed collision with other class~I nodes, we employ a mechanism based on the splitting algorithm \cite{DN} for resolving the collision. In particular, agent $k$ chooses a uniformly distributed integer, $C_k^{q_k}$, in the interval $[(q_k-1)m+1,q_km]$, where $q_k$ is node $k$'s collision counter and $m$  denotes the number of branches into which nodes involved in the latest collision are divided. As soon as the medium becomes idle, agent $k$ transmits a pulse for a duration of $C_k^{q_k}$ slots and then starts sensing the medium.  If it concludes that the following slot is busy, it defers its transmission and repeats this step, without changing its  collision counter. Otherwise, if the medium is idle after its pulse transmission, the agent starts its transmission. Hence, the node with the largest pulse transmission period among class~I nodes will be the winner of the contention.
\item If a collision happens during the collision resolution period, nodes involved in the collision increase their collision counters and go to the previous step.    

\item Each time a node transmits successfully, it resets its collision counter to~0 and follows the procedure for the Class~II agents.       
\end{enumerate}
\section{Fairness Analysis of DSCFQ}\label{fair}

Now let us introduce some definitions, which facilitate our fairness analysis of the proposed scheduling algorithm.

\begin{crp}\label{crpdef}
We define \emph{Collision Resolution Period}  as the whole period of the splitting algorithm, during which all class~I nodes receive service, and will refer to it as CRP in the rest of the paper. Furthermore, we define the set of nodes that were involved in the latest collision as \emph{Collision Set}.
\end{crp}

\begin{figure}[!t]
\centering
\noindent
\includegraphics[scale=0.7]{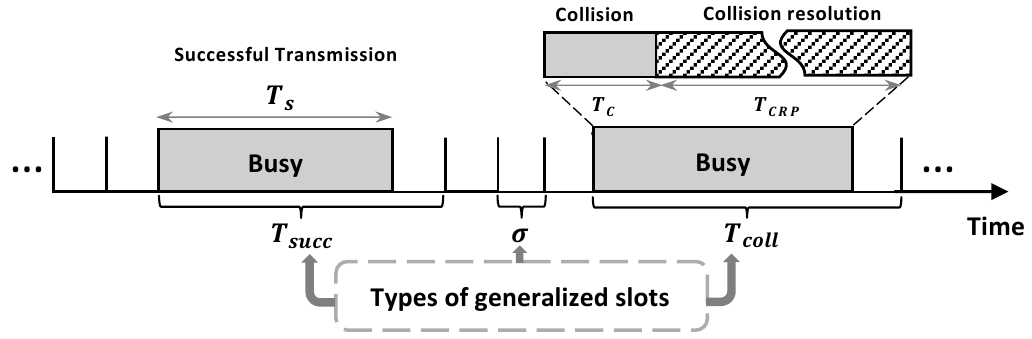}
\caption{Illustration of the possible states of a generalized time slot}
\label{fig:genslot}
\end{figure}

\begin{genslot}\label{genslot}
A generalized time slot is defined as the interval between two consecutive times at which class~II nodes are allowed to transmit a packet. 
\end{genslot}
\noindent In other words, the generalized time slot equals an idle slot~($\sigma$) if the medium has been idle at least for one time slot. Otherwise, 
the generalized time slot will be equal to the time interval during which the medium is entirely busy until it gets idle for one time slot. These cases are shown in Fig.~\ref{fig:genslot}.

\begin{virtual}\label{virtual}
We define the network's virtual time, $v(t)$, as the number of idle slots during interval $(0,t)$, normalized by the scaling factor. Furthermore, for any interval $(t_1,t_2)$, $v(t_1,t_2)$ is defined as $v(t_2)-v(t_1)$.
\end{virtual}
\noindent It is obvious from Definition~\ref{virtual} that the system's virtual time is a cumulative function and as a result, it is a nondecreasing function of time. Whenever a node transmits at the beginning of a slot, all the other nodes will be able to detect that transmission. As a result, all the agents have similar observations about the status of a given slot and therefore, all the agents are capable of tracking the network's virtual time.
 
In order to compare the normalized services received by different agents, we define virtual time and service deviation for each agent, as follows: 
 
\begin{flow_v}\label{flowv_def}
The virtual time of agent $k$, $v_k(t)$, is defined as
\begin{align*}
v_k(0)&=0,\qquad \forall k \in \mathcal{N},\\ 
v_k(t_1,t_2)&=\left \{ \begin{array}{ll}
w_k(t_1,t_2),      &\forall k \in \mathcal{B}(t_1,t_2),\\
v(t_1,t_2), &\forall k \in \mathcal{A}(t_1,t_2).
\end{array} \right. 
\end{align*} 
\end{flow_v}

\begin{diff}\label{servdev_def}
We define normalized service deviation of agent $k$, which represents the difference between the network's and the agent's virtual times, as follows
\begin{equation*}
\delta_k(t)=v(t)-v_k(t), \qquad \forall k \in \mathcal{N}.
\end{equation*}
\end{diff}

\begin{theo1}\label{maxdev}
The maximum normalized service disparity between any pair of agents $k$ and $j$, $k,j  \in ~ \mathcal{B}(t_1,t_2)$, in a shared medium with our proposed scheduling algorithm, is bounded as
\begin{equation}\label{fair_bound}
| w_k(t_1,t_2)-w_j(t_1,t_2) | \leq \frac{L_k^{max}}{\phi_k}+\frac{L_j^{max}}{\phi_j}+\frac{2}{\alpha},\end{equation}
where $L_k^{max}$ and $L_j^{max}$ are the maximum packet sizes that can be transmitted by agents $k$ and $j$, respectively.
\end{theo1}

We prove this theorem through the following set of lemmas. 
\begin{lem1}\label{fin_servdev}
    Whenever the transmission of an agent's message finishes, the normalized service deviation of the corresponding agent satisfies
    \begin{equation}\label{del_k}
     -\frac{1}{\alpha} < \delta_k(d_k^i) \leq 0 \qquad \forall i, \forall k\in\mathcal{N}.
    \end{equation}
\end{lem1}

\begin{proof}
Let us define $b_k^i$ as $\max\{a_k^i,d_k^{i-1}\}$. If packet $p_k^i$ arrives before that packet $p_k^{i-1}$ has finished its service, $b_k^i=d_k^{i-1}$ and hence
\begin{equation}\label{ee}
    \delta_k(b_k^i)=\delta_k(d_k^{i-1}), \qquad a_k^i<d_k^{i-1} . 
\end{equation}
Otherwise, $b_k^i=a_k^i>d_k^{i-1}$ and therefore we have
\begin{equation}\label{cc}
    \delta_k(b_k^i)=\delta_k(a_k^i)=\delta_k(d_k^{i-1})+\delta_k(d_k^{i-1},a_k^i).
\end{equation}
Since agent $k$ is not backlogged during the interval $(d_k^{i-1},a_k^i)$, we get $
\delta_k(d_k^{i-1},a_k^i)=v(d_k^{i-1},a_k^i)-v_k(d_k^{i-1},a_k^i)=0$.
We conclude from Eqs.~(\ref{ee}) and (\ref{cc}) that
\begin{equation}
\delta_k(b_k^i)=\delta_k(d_k^{i-1}) \label{16}.
\end{equation}
On the other hand, according to the definition of $b_k^i$,  all the previous messages of agent $k$ have been sent by time $b_k^i$. So, $p_k^i$ is  the only message that will be transmitted by agent $k$ during $(b_k^i,d_k^i)$. Furthermore, transmitting $p_k^i$ requires sensing the medium idle for $B_k^i$ generalized slots. Consequently, during the interval $(b_k^i,d_k^i)$, normalized service of agent $k$ and the network's virtual time will be increased by  $L_k^i/\phi_k$ and $B_k^i/\alpha$, respectively, i.e.,
\begin{align}
v_k(b_k^i,d_k^i)&=w_k(b_k^i,d_k^i)=\frac{L_k^i}{\phi_k}, \label{17}\\
v(b_k^i,d_k^i)&=\frac{B_k^i}{\alpha}=\frac{\lfloor \alpha (\frac{L_k^i}{\phi_k} - \epsilon_k^i ) \rfloor}{\alpha} \label{18}.
\end{align} 
Using Eqs.~(\ref{17}) and (\ref{18}), we get 
\begin{equation}
\delta_k(b_k^i,d_k^i)
=\frac{\lfloor \alpha (\frac{L_k^i}{\phi_k} - \epsilon_k^i ) \rfloor}{\alpha} - \frac{L_k^i}{\phi_k}. \label{21}
\end{equation}
Let us return to the definition of $\epsilon_k^i$. Comparing Eqs.~(\ref{eq:epsilon_def}) and (\ref{21}) we get $
\epsilon_k^{i+1} - \epsilon_k^{i}=\delta_k(b_k^i,d_k^i)$. 
Moreover, we have:
\begin{align*}
    \epsilon_k^i&=\epsilon_k^1+\sum_{n=1}^{i-1} (\epsilon_k^{n+1}-\epsilon_k^n)=\sum_{n=1}^{i-1} \delta_k(b_k^{n},d_k^n) \\
    &= \delta_k(b_k^{1},d_k^{1})+\sum_{n=2}^{i-1} \delta_k(b_k^{n},d_k^n).
\end{align*}
Using Eq.~(\ref{16}) and considering the fact that $\delta_k(b_k^1)=0$, since $b_k^1$ is the first time that agent $k$ gets backlogged and consequently $v(b_k^1)=v_k(b_k^1)$, we get the following equation:
\begin{equation}
    \epsilon_k^i=\delta_k(d_k^{1})+\sum_{n=2}^{i-1} \delta_k(d_k^{n-1},d_k^n) =\delta_k(d_k^{i-1}).\label{e_d}
\end{equation}
Therefore, \emph{CompensationFactor} for the $i$th message of agent $k$ is actually equal to the normalized service deviation after the transmission of message $p_k^{i-1}$. Using Eq.~(\ref{16}), we have $\delta_k(d_k^i)=\delta_k(d_k^{i-1})+\delta_k(b_k^i,d_k^i)$. Now,
defining $\gamma= \alpha ( \frac{L_k^i}{\phi_k}-\delta_k(d_k^{i-1}) )$, and substituting $\epsilon_k^i=\delta_k(d_k^{i-1})$ into Eq.~(\ref{21}), we get
\begin{equation}
    \delta_k(d_k^i)=\delta_k(d_k^{i-1})+\delta_k(b_k^i,d_k^i)=\frac{\lfloor \gamma \rfloor - \gamma}{\alpha}.
\end{equation}
Since $-1~<~\lfloor \gamma \rfloor - \gamma~\leq~0$, the proof is complete.
\end{proof}

\begin{lem2} \label{servdev}
The normalized service deviation of agent $k$, $k \in \mathcal{N}$, in a shared medium with our proposed scheduling algorithm, gets bounded as
\begin{equation}
-\frac{1}{\alpha} \leq \delta_k(t) \leq \frac{L_k^{max}}{\phi_k}, \qquad \forall k \in \mathcal{N}. \label{theorr}
\end{equation}
\end{lem2}
\begin{proof}
We consider two possible cases. In the first case, agent $k$ is absent at time $t$, i.e. $ k\in \mathcal{A}(t)$. If $p_k^{i-1}$ represents the last packet that was transmitted by this agent before $t$,  we can easily conclude that this agent's queue has been empty in the interval $(d_k^{i-1},t)$, which results in $v_k(d_k^{i-1},t)=v(d_k^{i-1},t)$, and therefore $\delta_k(d_k^{i-1},t)=0$.
Now, $\delta_k(t)$ can be calculated as follows:
\begin{align*}
\delta_k(t)&=\delta(d_k^{i-1})+\delta(d_k^{i-1},t)=\delta(d_k^{i-1}).
\end{align*}
Finally, using Lemma~\ref{fin_servdev} we get
\begin{equation}
-\frac{1}{\alpha} < \delta_k(t) \leq 0, \qquad k \in \mathcal{A}(t). \label{knotinB}
\end{equation}
Hence, while an agent is absent, its normalized service deviation equals $\delta(d_k^{i-1})$, where $p_k^{i-1}$ is the last packet sent by it before becoming absent.

Now, let us consider the other case in which agent $k$ is backlogged at $t$. Assuming that $p_k^{i}$ is the first packet that finishes its service after $t$, we get $b_k^{i} \leq t \leq d_k^{i}$, where $b_k^i=\max\{a_k^i,d_k^{i-1}\}$. So, we can calculate $\delta_k(t)$ as
$\delta_k(t)=\delta_k(b_k^{i})+\delta_k(b_k^{i},t)$.
Since the transmission of packet $p_k^i$ begins after sensing $B_k^i$ idle slots, the network's virtual time will have an increase of $B_k^i/\alpha$ by the time agent $k$ starts transmitting $p_k^i$, i.e.,  $t=s_k^i$. Furthermore, the medium will get busy during $(s_k^i,d_k^i)$ and therefore, the network's virtual time will remain unchanged during this interval. Therefore, for $t\in[b_k^i, d_k^i]$, $v(b_k^i,t)$ satisfies the following:
\begin{equation}
0 \leq v(b_k^i,t) \leq v(b_k^i,s_k^i)=\frac{B_k^i}{\alpha}=\frac{\lfloor \alpha ( \frac{L_k^i}{\phi_k}-\delta_k(d_k^{i-1}) ) \rfloor}{\alpha},  \label{virbnd}
\end{equation}
where we have used $\epsilon_k^i=\delta_k(d_k^{i-1})$ from Eq.~(\ref{e_d}).
On the other hand, agent $k$ is backlogged during $(b_k^i,d_k^i)$ and therefore, $v_k(b_k^i,t)=w_k(b_k^i,t)$. Since $s_k^i$ represents the time that transmission of packet $p_k^i$ begins, $w_k(b_k^i,t)$ will remain zero until $s_k^i$, and then reaches $L_k^i/\phi_k$ at time $d_k^i$ because of transmitting packet $p_k^i$ during $(s_k^i,d_k^i)$. So we have
\begin{equation}
0 \leq v_k(b_k^i,t)=w_k(b_k^i,t) \leq w_k(b_k^i,d_k^i) =\frac{L_k^i}{\phi_k}. \label{worbnd}
\end{equation}
Since $\delta_k(b_k^{i},t)=v(b_k^{i},t)-v_k(b_k^{i},t)$, from Eqs.~(\ref{virbnd}) and~(\ref{worbnd}) we deduce that $\delta_k(b_k^{i},t)$ is zero for $t=b_k^{i}$, reaches its maximum at $t=s_k^{i}$ and finally decreases to $\delta_k(b_k^i,d_k^i)$ at $d_k^i$. So, this yields 
\begin{equation*}
\min \left\{0,\delta_k(b_k^i,d_k^i)\right\} \leq \delta_k(b_k^i,t) \leq \delta_k(b_k^i,s_k^i).  
\end{equation*}
Adding $\delta_k(d_k^{i-1})$ to each side of the above inequalities, we get
 \begin{equation}
\begin{array}{ll}
\min \left\{\delta_k(d_k^{i-1}),\delta_k(d_k^i) \right\}  \leq \delta_k(t) \leq \delta_k(d_k^{i-1})+\delta_k(b_k^i,s_k^i), \label{ineq_delta} 
\end{array}
\end{equation}
where we have used Eq.~(\ref{16}).
Using Lemma~\ref{fin_servdev}, we have
\begin{equation}
-\frac{1}{\alpha} \leq \min \left\{\delta_k(d_k^{i-1}),\delta_k(d_k^i) \right\}.\label{lb}
\end{equation} 
Moreover, since $v_k(b_k^i,s_k^i) = w_k(b_k^i,s_k^i)=0$, the upper-bound in Eq.~(\ref{ineq_delta}) can be simplified as
\begin{align*}
&\delta_k(d_k^{i-1})+\delta_k(b_k^i,s_k^i) =\delta_k(d_k^{i-1})+v(b_k^i,s_k^i)  \\
&=\delta_k(d_k^{i-1})+\frac{\lfloor \alpha ( \frac{L_k^i}{\phi_k}-\delta_k(d_k^{i-1}) ) \rfloor}{\alpha}=\frac{\lfloor \gamma \rfloor - \gamma}{\alpha}+\frac{L_k^i}{\phi_k} \\ &\leq \frac{L_k^{max}}{\phi_k}.
\end{align*}
Therefore, using (\ref{ineq_delta}) and (\ref{lb}) we have
 \begin{equation}
-\frac{1}{\alpha} \leq \delta_k(t) \leq \frac{L_k^{max}}{\phi_k}, \qquad k \in \mathcal{B}(t), \label{kinB}
\end{equation}
 Finally, the lemma follows from (\ref{knotinB}) and (\ref{kinB}).
\end{proof}

It can be shown from Lemma~$\ref{servdev}$ that
\begin{equation}\label{bi_servdev}
| \delta_k(t_1,t_2) | \leq \frac{L_k^{max}}{\phi_k}+\frac{1}{\alpha}, \qquad k \in \mathcal{N}. 
\end{equation} 
Using Definitions~\ref{flowv_def} and~\ref{servdev_def}, and inequality~(\ref{bi_servdev}), we get the following corollary.


\begin{corr1}\label{corr}
For any agent $k \in \mathcal{B}(t_1,t_2)$, we have\\
\begin{equation}
| v(t_1,t_2)-w_k(t_1,t_2) | \leq \frac{L_k^{max}}{\phi_k}+\frac{1}{\alpha}.
\end{equation}

\end{corr1}

Now, Theorem~\ref{maxdev} can be easily obtained by using Corollary~\ref{corr} for any pair of agents $k$ and $j$, $k,j  \in ~ \mathcal{B}(t_1,t_2)$.

It can be observed from Theorem~\ref{maxdev} that the maximum normalized service disparity between a pair of backlogged agents is a function of the maximum packet sizes, fairness weights and the scaling factor. Furthermore, the effect of discretizing backoff times after scaling by $\alpha$ is reflected in the term  $2/\alpha$ in Eq.~(\ref{fair_bound}). Hence, using small scaling factors results in large  maximum service deviations and vice versa. 
\section{Throughput Analysis of DSCFQ}\label{throu}
In this section, we study the achievable network throughput by our proposed algorithm. We begin by introducing some assumptions and parameters that will be used in the  analysis. 

 We concentrate on saturation throughput, which represents the network throughput in overloaded conditions that all the agents are always backlogged.
Let us denote by $g_k$ the probability that agent $k \in \text{class II}$ attempts to transmit a packet in a given generalized slot. Furthermore, we define total attempt rate $G$ as the expected number of  transmission attempts in a generalized slot, i.e., $G=\sum_{n=1}^N g_k$. 
Now, we analyze the network throughput for the case in which $N\gg1$, and assume that each agent's average access probability is very small compared to the total transmission attempt rate, i.e., $g_k \ll G, \forall k \in \mathcal{N}$.
Since $G$ represents the average  number of transmission attempts in a given generalized slot, and also each collision wastes much more time compared to an idle slot, we expect the optimal total attempt rate, which maximizes the network saturation throughput, to be less than one. So, we conduct the analysis for the case in which $G<1$, and consequently  $g_k \ll 1, \forall k \in \mathcal{N}$.


 As discussed in the previous section, a generalized time slot  may be an idle slot, may contain a successful transmission or it may contain collisions. So, a generalized time slot  might have a length of $T_{idle}=\sigma$, $T_{succ}$ or $T_{coll}$ as shown in Fig.~\ref{fig:genslot}.  Let us denote by $T_s$ and $T_c$ the duration of a successful transmission and the length of the wasted time caused by the collision of class~II agents, respectively. Hence,  $T_{coll}$ and $T_{succ}$ can be calculated based on Fig.~\ref{fig:genslot} as follows:
 \begin{equation*}
T_{coll}=T_c+T_{CRP}+\sigma,\qquad
T_{succ}=T_s+\sigma,
\end{equation*} 
where $T_{CRP}$ represents the duration of the contention resolution period (CRP).
Now, we calculate the probabilities that a given generalized slot is idle ($P_{idle}$), it contains a successful transmission ($P_{succ}$) or it contains collisions ($P_{coll}$). Since each agent $k\in\text{class II}$, transmits with probability $g_k$ in a generalized slot, the total number of transmission attempts in a given slot has a Poisson Binomial distribution. Furthermore, Le Cam's theorem \cite{lecam} suggests that the Poisson Binomial distribution with $g_k~\ll~1, \forall k \in \mathcal{N}$, can be approximated by a Poisson distribution with mean $G=\sum_{k=1}^{N} g_k$. Therefore, considering the fact that a given slot remains idle (contains a successful transmission) only if zero (one) transmission attempt has been made in that slot, we can calculate $P_{idle}$, $P_{succ}$ and $P_{coll}$ as follows
\begin{align}\label{Probs}
P_{idle}&\simeq  e^{-G}, \qquad P_{succ}\simeq G e^{-G}, \nonumber \\
P_{coll}&=1-P_{succ}-P_{idle}=1-G e^{-G}-e^{-G}.
\end{align}
 Now, let us define the normalized saturation throughput, which is represented by $S$, as the fraction of time  the medium is used to successfully transmit payload bits. We can express $S$ as follows:
\begin{equation}
S=\frac{(P_s+\bar{n}_c P_c)(\bar L/C)}{P_s T_{succ} + P_i \sigma+ P_c T_{coll}}, \label{throughput}
\end{equation}
where $\bar{L}$, $C$ and $\bar{n}_c$ denote the average message length, the transmission rate, and the average number of agents involved in a given collision of class~II nodes, respectively. Let $n_t$ denote the number of agents attempting a transmission in a given generalized slot. Since a collision happens only if the number of transmitted packets in a given time slot is larger than one, $\bar{n}_c$ can be calculated as follows
\begin{equation}
\bar{n}_c = E[n_t|\textnormal{collision}]=E[n_t|n_t>1]
=G (1-e^{-G}). 
\end{equation}

\section{Adaptive DSCFQ}\label{adapt_dscfq}
As we discussed earlier, scaling factor is a parameter for backoff adjustment and it has a direct impact on the network throughput. Choosing a small scaling factor results in an increase of the collision probability, and reduces the idle intervals during which the medium is left unused. On the other hand, increasing the scaling factor causes less collisions, but large idle intervals. Therefore, it is important to pick the optimal scaling factor that maximizes the throughput. In the following, we propose a simple adaptive method for updating the scaling factor to reach the peak saturation throughput.

Let us first define the optimal total attempt rate, $G^*$, as the value for which the saturation throughput in Eq.~(\ref{throughput}) is maximized.  Substituting $G^*$ in Eq. (\ref{Probs}), we get $P^*_{idle}$, $P^*_{succ}$ and $P^*_{coll}$, respectively.
Now, we introduce an adaptive method to reach the maximum saturation throughput. As discussed earlier, each generalized slot has one of the three different states of successful transmission, collision or idle state. Assume that all the agents agree on a common initial scaling factor, which is denoted by $\alpha_0$. Then each agent senses the medium and updates $\alpha$ based on each generalized slot's \textit{state}  as follows:
\begin{equation}
\alpha_{new} = \left\{
\begin{array}{ll}
\alpha ,  \qquad &\textnormal{success}\\ 
\alpha + \gamma,  \qquad &\textnormal{collision}\\ 
\alpha - \beta, \qquad &\textnormal{idle} 
\end{array} \right. \label{alpha_update}
\end{equation}
where $P_{idle}^* \beta=P_{coll}^* \gamma.$

Now, let $\alpha^*$ denote the optimal scaling factor that maximizes the saturation throughput. Since  $P_{idle}$ and $P_{coll}$ are, respectively, monotonically decreasing and  increasing functions of $G$, and $G$ is inversely proportional to $\alpha$, we have
\begin{equation}
D_{\alpha}=\left\{
\begin{array}{ll}
\gamma P_{coll} - \beta P_{idle} > 0,  \qquad & \textnormal{if $\alpha<\alpha^*$}\\ 
\gamma P_{coll} - \beta P_{idle} < 0,  \qquad & \textnormal{if $\alpha>\alpha^*$}
\end{array} \right. \label{drift}
\end{equation}
where $D_{\alpha}$ is the expected drift of the scaling factor from one generalized slot to the next one. Hence, as long as $\alpha < \alpha^*$, there will be a positive drift which increases $\alpha$ on average. On the other hand, $\alpha > \alpha^*$ results in a negative drift and therefore, leads to the reduction of the scaling factor.

\section{Performance Evaluation}\label{numerics}
In this section, we present our evaluations of the proposed distributed scheduling algorithm. We first describe the experimental setup and the technical assumptions used in the experiments. We then explain the baselines and the metrics that have been used for the evaluations. Finally, we discuss the experiments and results.
 
\begin{table}[!t]
\centering
\caption{\\Simulation Parameters}
\begin{tabular}[!t]{l|r@{ }l|l}
\hline
\rowcolor{Gray}
Parameter & \multicolumn{2}{c}{Value} & Definition \\
\hline
\hline
$C$ & 12 &Mbps & data transmission rate \\ 
$C_{ctrl}$ & 6 &Mbps & control transmission rate\\
\hline \hline
$L_{ACK}$ & 112 &bits & ACK frame length\\ 
$L_{RTS}$ & 160 &bits & RTS frame length\\ 
$L_{CTS}$ & 112 &bits &CTS frame length\\  
$\bar L$ & 2016 &bytes & average message length\\
\hline \hline
$\sigma$ & 9 &$\mu$s & slot time\\  
$SIFS$ & 10 &$\mu$s & SIFS time\\  
\hline
\end{tabular}
\end{table}

\begin{figure*}[!t] 
\centering
\subcaptionbox{w=30}{\includegraphics[width=0.22\linewidth, height=2.7cm]{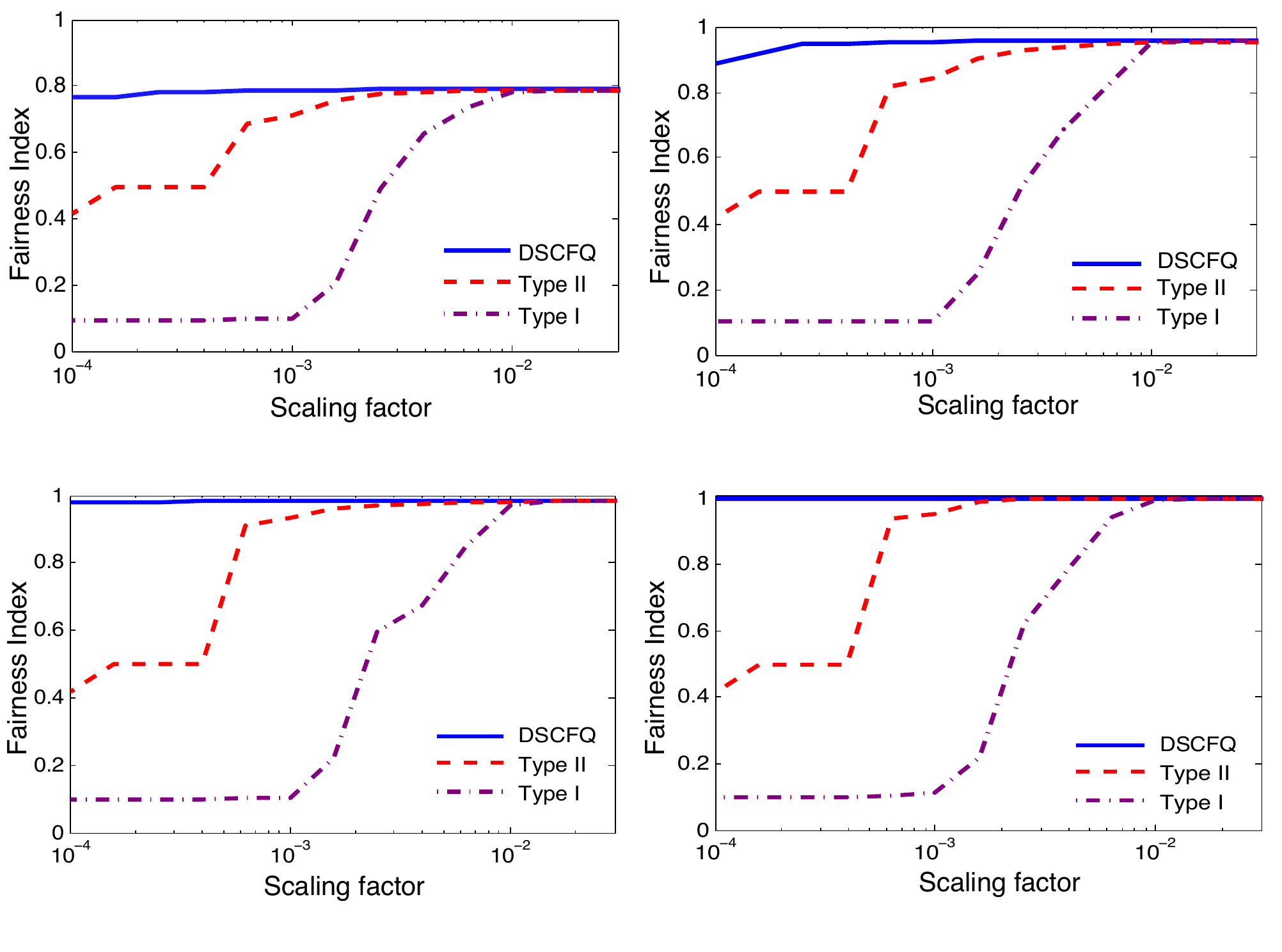}}
\subcaptionbox{w=50}{\includegraphics[width=.22\linewidth, height=2.7cm]{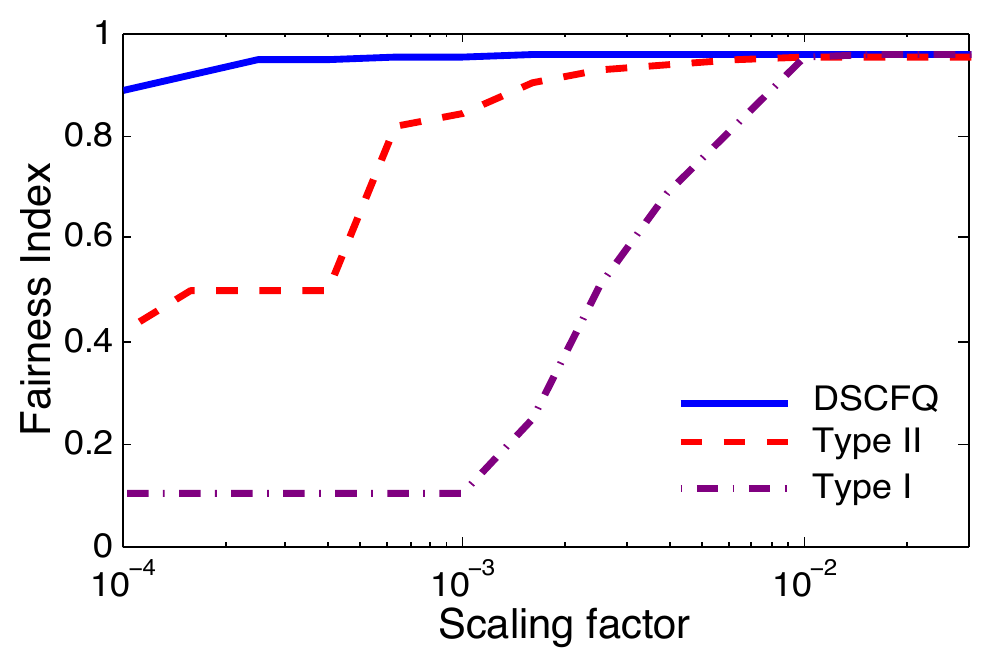}}
\subcaptionbox{w=100}{\includegraphics[width=.22\linewidth, height=2.7cm]{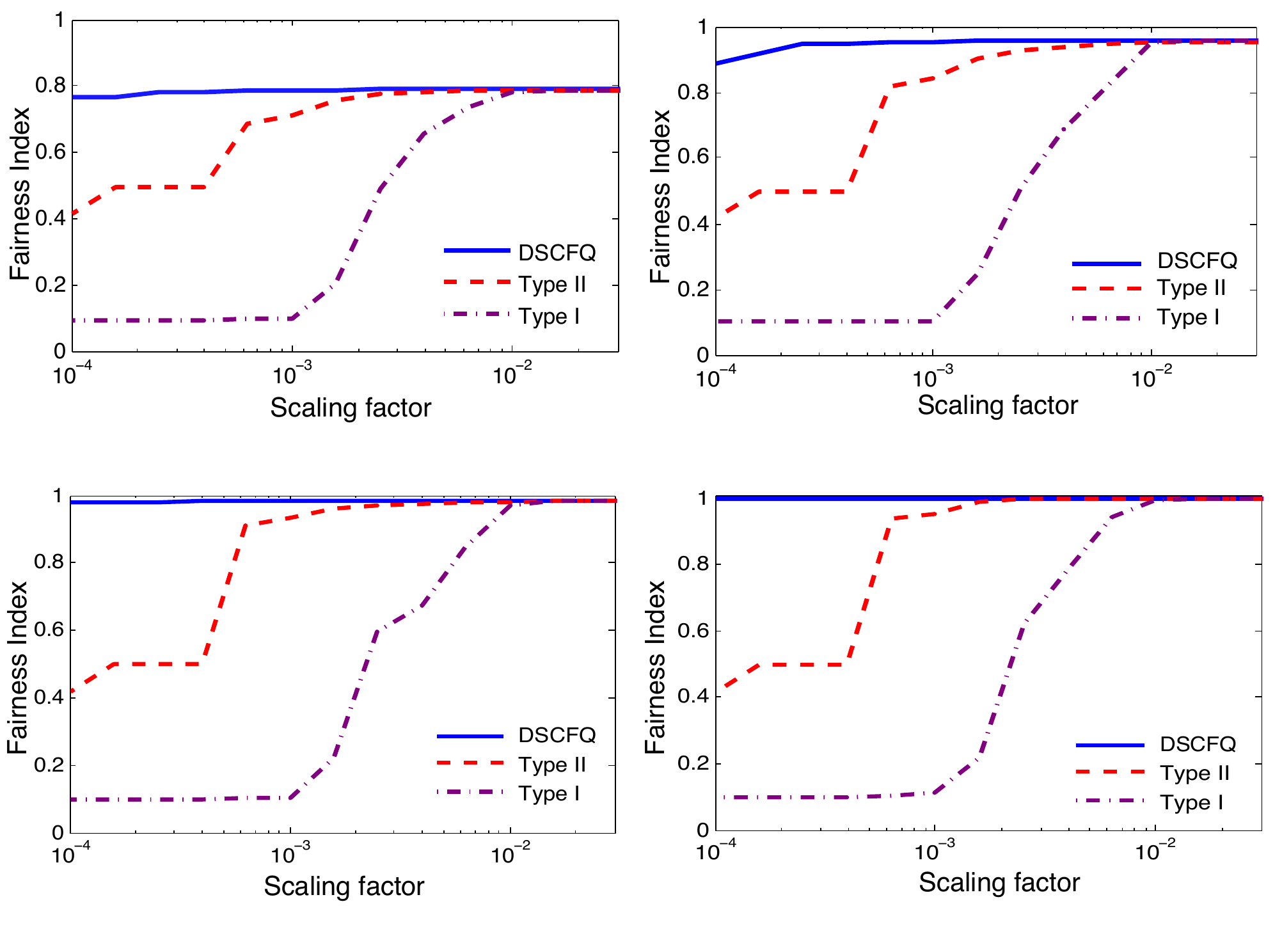}}
\subcaptionbox{w=1000}{\includegraphics[width=.22\linewidth, height=2.7cm]{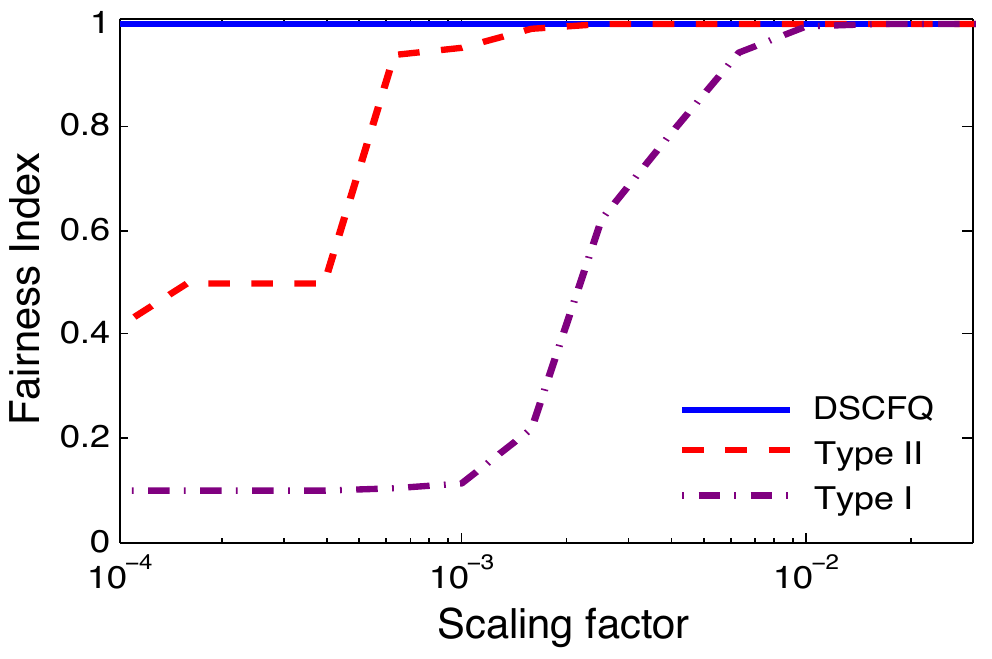}}
\caption{Average fairness index as a function of the scaling factor, for window sizes $w$=30, 50, 100, 1000.}
\label{fig:stermfairness_win_tot}
\end{figure*}

\textbf{Experimental Setup:}
Consider a network of 10 agents that are always backlogged and need to share their information with each other. Furthermore, the importance of the agents' information are different. Therefore, each agent chooses a fairness weight that reflects the importance of its messages. We consider a scenario in which three agents choose  $\phi=10$, three other agents choose $\phi=8$, two agents pick $\phi=2$ and finally two agents pick $\phi=1$ as their fairness weights. Moreover, to study the performance of our algorithm in more realistic conditions, we implement our algorithm based on 802.11 DCF. Specifically, in our implementation we consider various practical requirements such as the control messages (RTS/CTS/ACK), inter frame spacing intervals, the imperfections caused by the packet headers and the propagation delay, etc.
The parameters used in our implementation are summarized in Table~1. 

\textbf{Baselines:}
To compare our algorithm with the existing fair scheduling algorithms, we classify these baselines into two general types. The first type, which we call Type~I, includes algorithms that emulate weighted fair queueing by taking backoff intervals (or IFS values) in proportion to the packet size of the head of the queue messages, i.e. $B_k^i=\lfloor \alpha (\frac{L_k^i}{\phi_k}) \rfloor$, while still using random mechanisms such as BEB for collision resolution. The DFS algorithm introduced by~\cite{DFS} belongs to this category.  The other type (Type~II) uses similar methods for providing fairness, with the difference of giving a higher priority to the agents with unsuccessful last transmission attempts. The proposed algorithm by \cite{pdfs} is an example of type~II algorithms. An advantage of the above classification is that the value and impact of each part of our algorithm on the performance of the system can be assessed separately.

\begin{figure}[!t] 
\centering
\subcaptionbox{}{\includegraphics[width=0.32\linewidth, height=2.7cm]{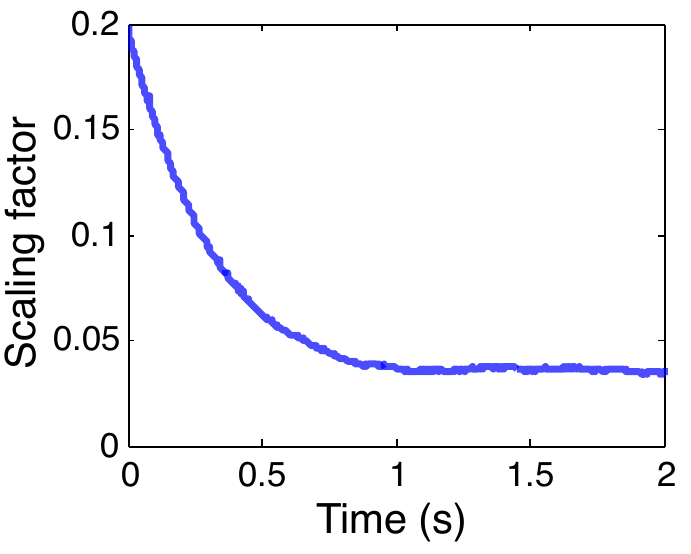}}
\subcaptionbox{}{\includegraphics[width=.32\linewidth, height=2.7cm]{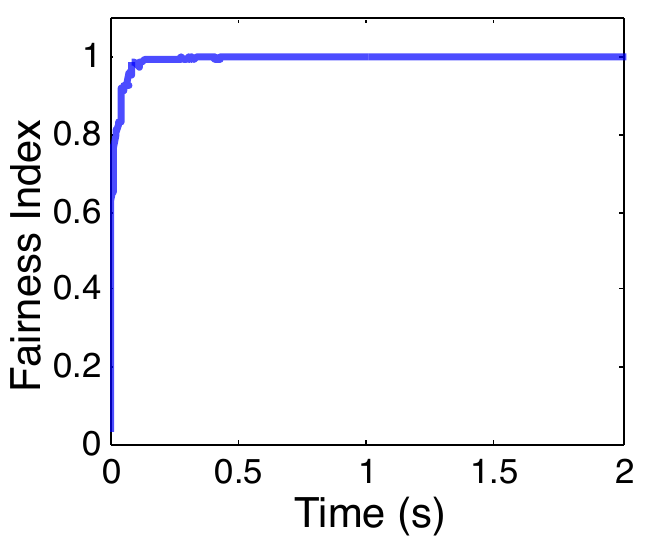}}
\subcaptionbox{}{\includegraphics[width=.32\linewidth, height=2.7cm]{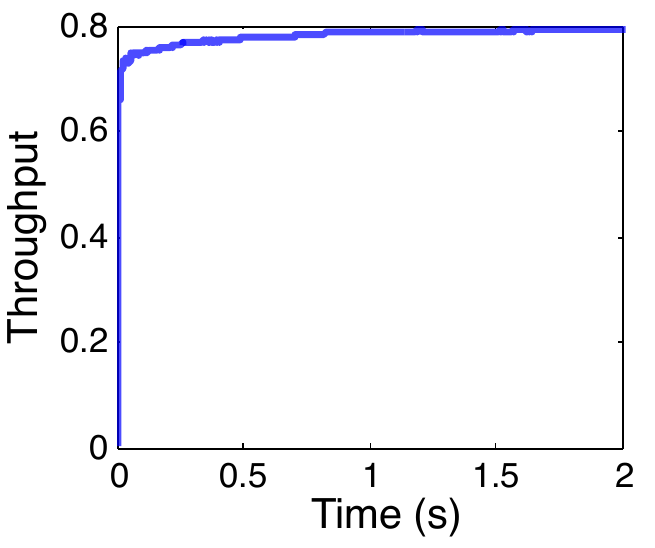}}
\caption{Convergence of the (a) scaling factor (b) fairness index and (c) normalized throughput to their optimal values. }
\label{fig:running_DSCFQ}
\end{figure}

\textbf{Evaluation Metrics:}
For evaluating the fairness of our algorithm, we use the \textit{fairness index} \cite{findex}, which is defined as follows:
\begin{equation}
\text{fairness index}=\frac{ \big( \sum_{k} T_k/\phi_k \big)^2}{N \cdot \sum_{k} \big(T_k/\phi_k \big)^2  },
\end{equation}
where $T_k$ denotes the throughput of flow $k$. Moreover, we compare the degree of fairness over different time-scales using Sliding Window Method (SWM), introduced by \cite{short_fair}. In this method, we slide a window of size $w$ across a packet trace of medium accesses. By calculating the fairness index for the transmissions that lie in this window after each sliding, we get a sequence of fairness indices. We report the average fairness index as the fairness metric for a given window size. 
The other metrics used in our evaluations are the throughput of the agents and the network throughput. The throughput of a particular agent in a given interval is defined as the ratio of the successfully transferred information to the length of the interval. The network throughput can be calculated by adding the throughput of the agents.

\subsection*{Results and Discussions}
Let us start our evaluations with the fairness analysis of the system. Fig. \ref{fig:stermfairness_win_tot} shows the performance of our algorithm along with type~I and~II scheduling schemes for different time scales ($w$=30, 50, 100 and 1000). As can be observed, our algorithm achieves the best performance among all the compared algorithms in both short-term ($w$=30, 50) and long-term ($w$=100, 1000) comparisons. While our algorithm achieves a relatively constant fairness index over a wide range of scaling factors ([0.0001,0.02]), the performance of type I and II algorithms deteriorate dramatically for the small scaling factors. This is particularly important since large scaling factors can result in more idle time slots and therefore, less efficient utilization of the medium. On the other hand, we can observe that the difference between the  fairness indices of the compared algorithms starts to diminish as we increase the scaling factor. This can be explained by considering the main sources of unfairness in Type~I and~II algorithms. Specifically, the agents with unsuccessful last transmissions have no priority over the other agents in type~I algorithms, which creates unfairness in the case of collisions. Since increasing the scaling factor results in less collisions, the effect of this factor becomes less dominant. On the other hand, discretization of the backoff intervals, i.e., using $\lfloor \alpha (L_k^i /\phi_k) \rfloor$ instead of $\alpha (L_k^i/\phi_k)$, is another source of unfairness in both type~I and II algorithms. In a similar way, the impact of this factor is more noticeable for small scaling factors, where a large range of packet lengths can be mapped into the same backoff intervals after scaling and discretization. The compensation factor $\epsilon_k^i$ used in the backoff calculation step in our algorithm (Eq.~(\ref{bkoff})) deals with this issue effectively. Since our algorithm eliminates the effect of the two sources of unfairness, it achieves the best performance among all the compared baselines in different time scales (Fig.~\ref{fig:stermfairness_win_tot}). Therefore, the difference between the fairness indices of type I and II algorithms shows the amount of improvement in average fairness index, using an algorithm that gives priority to the agents with unsuccessful last transmission attempts. On the other hand, the difference between our algorithm and type II class demonstrates the increase in the fairness index by using compensation factor in the backoff calculation as in Eq.~(\ref{bkoff}).
Another observation we make is that the fairness index generally decreases for all the algorithms as we consider shorter time scales. This phenomenon can be also observed in the centralized scheduling algorithms, which is justified by the packet-based nature of the problem (in contrast to fluid flow models).

 

\begin{figure}[!t] 
\centering
\subcaptionbox{}{\includegraphics[width=0.45\linewidth, height=2.7cm]{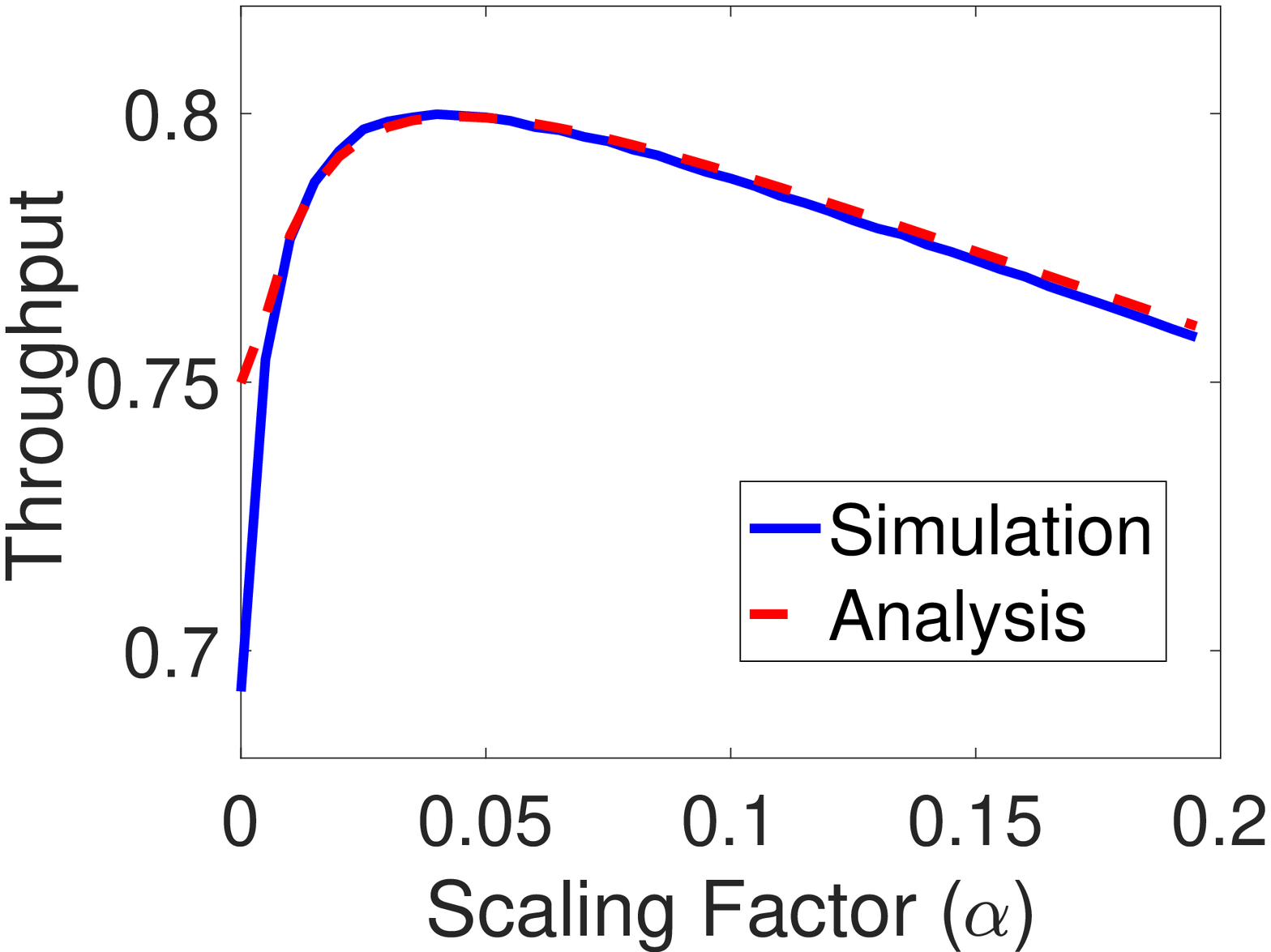}} 
\subcaptionbox{}{\includegraphics[width=0.45\linewidth, height=2.7cm]{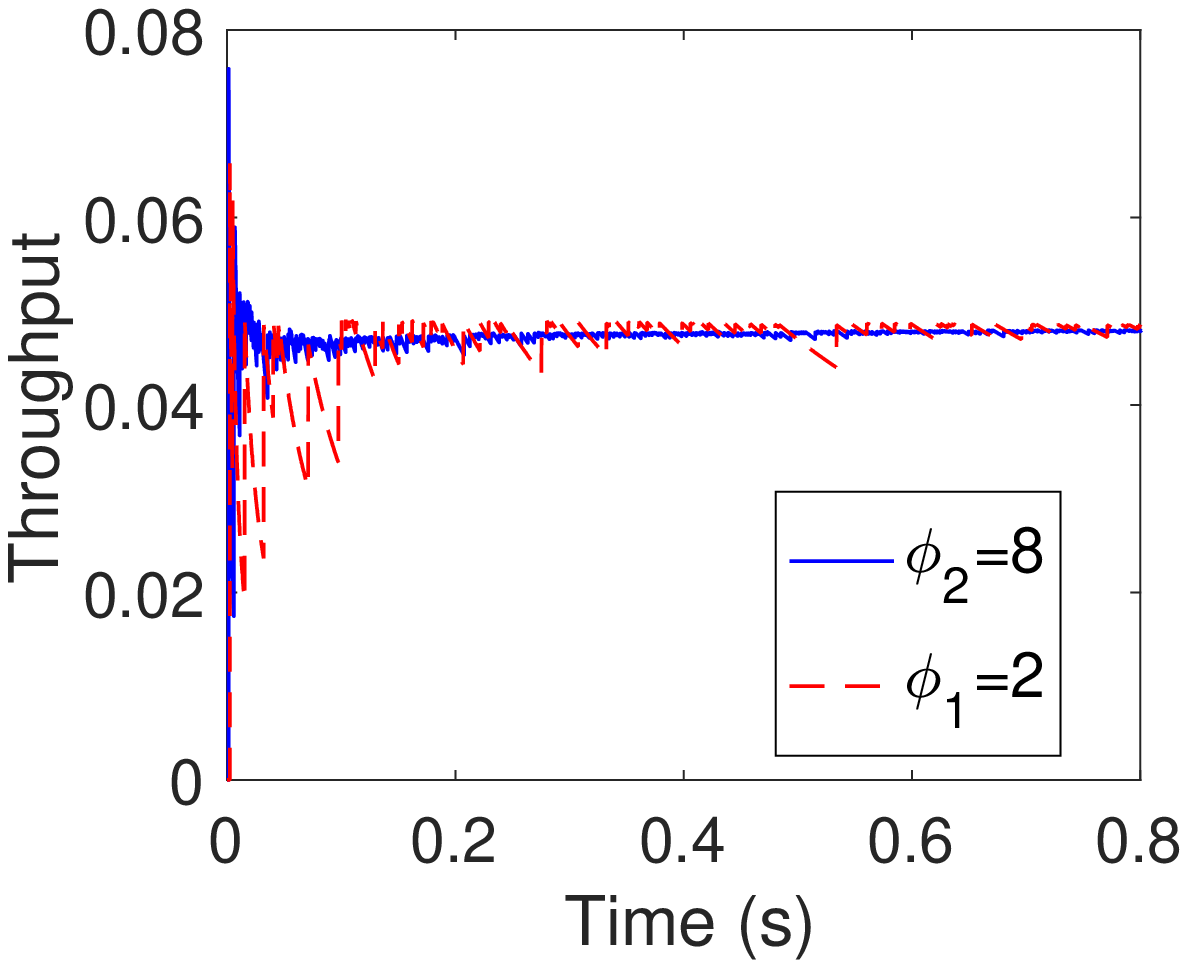}} 
\caption{a) Comparison of the normalized network throughput obtained from the simulation and Eq.~(\ref{throughput}), as a function of the scaling factor. b) Convergence of the normalized throughput of agents 1 and 2 (with $\phi=2$ and $\phi=8$) into the same shares}
\label{fig:throughput_alpha_time}
\end{figure}

Now, let us study the adaptive behaviour of our algorithm. As we discussed before, there is an intrinsic trade-off between the fairness and the network throughput, which can be controlled by the scaling factor. In previous section, we proposed an adaptive method for updating the scaling factor to achieve the maximum throughput attainable by our algorithm. Fig.~\ref{fig:running_DSCFQ}a shows dynamic adaptation of the scaling factor for the case in which it is initially set to 0.2. We observe that as the scaling factor  converges into its optimum value, the fairness index approaches one (Fig.~\ref{fig:running_DSCFQ}b) and the network throughput reaches its maximum value (Fig.~\ref{fig:running_DSCFQ}c). This can be validated from Fig.~\ref{fig:throughput_alpha_time}a, which shows the normalized network throughput obtained from the simulation and the theory (Eq.~(\ref{throughput})). As can be seen, the analytically driven throughput from Eq.~(\ref{throughput}) closely approximates the throughput obtained from the simulations. Moreover, we can observe that the maximum throughput ($\simeq 0.8$) is achieved for a scaling factor around $0.04$, which are the same values that our algorithm converges into in Figs.~\ref{fig:running_DSCFQ}a and \ref{fig:running_DSCFQ}c.
Finally, Fig.~\ref{fig:throughput_alpha_time}b shows the convergence of the normalized throughput of two agents with fairness weights of $\phi_1=2$ and $\phi_2=8$. As can be observed, our algorithm equalizes the normalized throughput of these agents after its convergence.

\section{Conclusion}\label{conclusion}
In this paper, we considered one of the important, but understudied challenges in the communication of multi-agent systems. Specifically, we studied the shared medium contention problem in the communication of the agents and proposed a new distributed scheduling algorithm for providing fairness in these systems. This becomes even more important when the agents' information or observations have different importance, in which case the agents require different priorities for accessing the medium and sharing their information. We showed that our proposed algorithm can provide a deterministic bound on the maximum service disparity among any pair of agents. This can particularly improve the short-term fairness, which is important in real-time applications. Moreover, we designed an adaptive mechanism that enables our scheduling algorithm to adjust itself and achieve a high throughput at the same time.

\bibliography{reference}

\end{document}